# RPA natural orbitals and their application to post-Hartree-Fock electronic structure methods

View Online    Export Citation    CrossMark

Benjamin Ramberger,[1] Zoran Sukurma,[1] Tobias Schäfer,[2] and Georg Kresse[1,a]

**AFFILIATIONS**

[1] Faculty of Physics and Center for Computational Materials Sciences, University of Vienna, Sensengasse 8/12, 1090 Vienna, Austria
[2] Institute for Theoretical Physics, Technical University of Vienna, Wiedner Hauptstraße 8-10/136, 1040 Vienna, Austria

[a] Electronic mail: georg.kresse@univie.ac.at

## ABSTRACT

We present a method to approximate post-Hartree-Fock correlation energies by using approximate natural orbitals obtained by the random phase approximation (RPA). We demonstrate the method by applying it to the helium atom, the hydrogen and fluorine molecule, and to diamond as an example of a periodic system. For these benchmark systems, we show that RPA natural orbitals converge the MP2 correlation energy rapidly. Additionally, we calculated full configuration interaction energies for He and $H_2$, which are in excellent agreement with the literature and experimental values. We conclude that the proposed method may serve as a compromise to reach good approximations to correlation energies at moderate computational cost, and we expect the method to be especially useful for theoretical studies on surface chemistry by providing an efficient basis to correlated wave function based methods.

## I. INTRODUCTION

In *ab initio* quantum chemistry and computational materials physics, there exists a well-known trade-off between accuracy and computational cost. Solving the many electron Schrödinger equation is an exponentially hard problem with respect to the number of electrons, and therefore, solving it exactly is out of scope for most practically relevant systems. Therefore, approximations with varying degrees of accuracy are employed. While mean field methods like Hartree-Fock (HF)[1–3] and density functional theory (DFT)[4,5] possess computationally favorable scaling of $N^3$ with the number of electrons, they sometimes lack accuracy and fail to describe certain processes. For example, describing the dissociation of simple molecules like $H_2$ already poses a challenge for such methods.[6] The properties that are not well described by the HF approximation are attributed to the so-called electron correlation, and there exists a wide variety of electronic structure methods all of which attempt to take the correlation effects into account in an approximate manner. Most of them have in common that they can increase the accuracy systematically at the expense of computational resources, e.g., Møller-Plesset perturbation theory (MPn),[7,8] coupled cluster (CC),[9–11] and configuration interaction (CI)[12–18] methods.

DFT and HF calculations can provide a useful single electron basis as a starting point for correlated methods. However, the correlated methods have in common that they usually involve virtual orbitals, i.e., eigenstates of the single-electron Hamiltonian from the underlying HF or DFT calculation that are unoccupied in the respective underlying calculation. To obtain the energy exactly (within the approximation of the respective correlated method), a complete basis set of single particle states is required. The convergence behavior with respect to the number of single particle states depends of course on the specific set of orbitals employed. Therefore, one can in principle obtain results that are converged up to machine precision with a finite number of virtual states and ideally reduce the computational costs significantly by using a small number of appropriate single particle orbitals.

The essence of the above statement is that by choosing an appropriate set of single particle orbitals, one can extract the relevant information resting in the full many body wave function from a small number of single particle wave functions. The question remains how to choose an appropriate set. One possible answer is to use natural orbitals as they turned out to capture much of the relevant information required to converge correlation energies quickly.[19] It has already been demonstrated that natural orbitals at the MP2 level can be employed to reduce the computational cost of wave function based methods.[20] Furthermore, recently Bruneval had shown that the so-called linearized GW density matrix is of similar quality to the MP2 density matrix.[21,22] This and the connection[23] between the random phase approximation (RPA)[24,25] and GW method,[26] as well as the successful applications of the RPA,[27–31] suggests that RPA natural orbitals (RPANOs) should provide an efficient basis for higher hierarchy methods.

In this article, we will show how we can approximate natural orbitals by employing the RPA and how to utilize the approximate natural orbitals to obtain accurate correlation energies at moderate computational cost. We will demonstrate the method by showing the convergence behavior of the MP2 energy with respect to the number of RPANOs for the following prototypical systems: the helium atom, the hydrogen molecule at various internuclear distances, the fluorine molecule at equilibrium distance, and diamond as an example for a periodic system. In the case of He and $H_2$, we will also show results for full configuration interaction method (FCI) calculations using an RPANO basis and will furthermore compare the exact natural orbitals from FCI with the RPANOs.

## II. THEORY

In second quantization notation, the one particle reduced density matrix (1-RDM) is

$$n(\mathbf{r}, \mathbf{r}') = \langle \Psi | \hat{\psi}^\dagger(\mathbf{r}') \hat{\psi}(\mathbf{r}) | \Psi \rangle, \quad (1)$$

with the usual field operator $\hat{\psi}$. Natural orbitals $\{\varphi_i\}$ are just the eigenstates of the 1-RDM,

$$\int_{-\infty}^{\infty} d^3\mathbf{r}' \, n(\mathbf{r}, \mathbf{r}') \, \varphi_i(\mathbf{r}') = f_i \, \varphi_i(\mathbf{r}). \quad (2)$$

The 1-RDM is the central object of reduced density matrix functional theory (RDMFT).[32] In comparison with conventional DFT, RDMFT has the advantage that besides the Hartree energy, also the kinetic and the exchange energy are obtained exactly. Though an efficient way to approximate the 1-RDM as well as natural orbitals can therefore provide an interesting pathway for RDMFT.[33] Though we did not pursue it, we still want to note that the method described hereafter for approximating the 1-RDM within the RPA might also be interesting for research in the field of RDMFT.

We will now try to establish a connection between the 1-RDM and the RPA by using a Green's function formalism. The 1-RDM is obtained from the Green's function in the limit of small negative time,

$$n(\mathbf{r}, \mathbf{r}') = -i \lim_{t \to 0^-} G(\mathbf{r}, \mathbf{r}', t), \quad (3)$$

which can be easily seen from the definition of the Green's function,

$$G(\mathbf{r}, \mathbf{r}', t' - t) = -i\langle \Psi_0 | \hat{T}\{\hat{\psi}(\mathbf{r}, t)\hat{\psi}^\dagger(\mathbf{r}', t')\} | \Psi_0 \rangle, \quad (4)$$

where $\hat{T}$ is the Wick time-ordering operator.

In a system with $N$ electrons, the independent particle Green's function $G^0$ corresponds to a ground state with a single Slater determinant $|\phi_1 \cdots \phi_N\rangle$ and the 1-RDM reduces to

$$n^0(\mathbf{r}, \mathbf{r}') = -i \lim_{t \to 0^-} G^0(\mathbf{r}, \mathbf{r}', t) = \sum_{i \in occ.} \phi_i(\mathbf{r})\phi_i^*(\mathbf{r}'). \quad (5)$$

The natural orbitals are then just the occupied independent particle orbitals $\{\phi_i\}$. This would be the situation for a DFT or HF calculation.

To improve on the independent particle 1-RDM and to include "information" on the unoccupied virtual space, the Dyson equation for the Green's function

$$G(2, 1) = G^0(2, 1) + G^0(2, 3)\Sigma(3, 4)G(4, 1) \quad (6)$$

can be used to write a perturbation series for the (full) 1-RDM as a functional of the total self-energy $\Sigma$ and the independent particle Green's function $G^0$,

$$n = n^0 - i \lim_{t \to 0^-} \left\{ G^0\Sigma G^0 + G^0\Sigma G^0\Sigma G^0 + \ldots \right\}. \quad (7)$$

In the imaginary time/frequency domain, the first order term (with respect to $\Sigma$), $n^{(1)}$, is explicitly obtained as[34–37]

$$n^{(1)}(\mathbf{r}, \mathbf{r}') = \lim_{\tau \to 0^-} \frac{1}{2\pi} \int_{-\infty}^{\infty} d\nu \, e^{-i\nu\tau}$$
$$\times \int d\mathbf{r}'' d\mathbf{r}''' G^0(\mathbf{r}, \mathbf{r}'', \nu)\Sigma(\mathbf{r}'', \mathbf{r}''', \nu)G^0(\mathbf{r}''', \mathbf{r}', \nu). \quad (8)$$

In the method shown below, we will use this first order correction to the density matrix. Furthermore, we will approximate the self-energy using the RPA so that we will ultimately obtain an approximate set of natural orbitals, the RPANOs, as eigenfunctions of the first order RPA density matrix,

$$n^{RPA}(\mathbf{r}, \mathbf{r}') = -i \lim_{\tau \to 0^-} \left[ G^0 + G^0\Sigma^{RPA}G^0 \right](\mathbf{r}, \mathbf{r}', \tau). \quad (9)$$

We note that the RPA violates the Pauli principle since it does not include all properly antisymmetrized diagrams (see, e.g., Kosov[38] or Hummel and co-workers[39]). Here, it is important to recall that the main purpose of our method is an efficient low scaling calculation of an approximate density matrix to allow one to obtain a compact set of orbitals for wave function based calculations. A highly accurate description of the density matrix is not strictly required in the present case. We will document one case where the RPA density matrix is qualitatively wrong but still yields a very compact set of orbitals (compare Fig. 13). More details on the RPA density matrix can be found in Refs. 21–23, 34, and 36.

## III. METHOD

Assuming that with only a few RPA natural orbitals one can span the relevant subspace of the one electron wave functions for advanced methods like MP2 or CI, we aim to accurately approximate electronic ground state energies of these methods using a small number of virtual states. Thereby, we are hopefully able to reduce the computational cost in these calculations significantly. In Subsection III A, we will give a step by step recipe for obtaining RPA natural orbitals in order to calculate approximate post-Hartree-Fock ground state energies (e.g., MP2, CC, and CI) with a reduced number of virtual states.

## A. RPA natural orbitals

In the first step, one calculates the electronic ground state at mean field level (HF or DFT) in order to obtain a well converged set of one electron wave functions for the occupied space. These occupied orbitals fix the respective mean field Hamiltonian for the subsequent steps. In the second step, "all" unoccupied orbitals are calculated by diagonalizing the one particle Hamiltonian in the entire underlying basis set. "All" means that the number of eigenfunctions of the Hamiltonian (i.e., orbitals) that are calculated is equal to the number of basis functions used. Specifically for a plane wave basis with a cutoff energy $E_{cut}$, this means that the sum of occupied orbitals $N_{occ}$ and virtuals $N_{virt}$ is equal to the number of plane waves below the cutoff, or in other words equal to the size of the underlying plane-wave basis set,

$$N_{occ} + N_{virt} = \mathrm{card}\left\{ \mathbf{G} : \frac{\hbar^2}{2m_e}|\mathbf{G}+\mathbf{k}|^2 < E_{cut} \right\}, \quad (10)$$

where $\mathbf{k}$ is the Bloch wave vector.

In the next step, the independent particle Green's function $G^0$ is calculated from all orbitals to subsequently calculate the independent particle polarizability $\chi^0$, the RPA screened interaction $W^{RPA}$, as well as the RPA self-energy $\Sigma^{RPA}$.[40–42] Note that $\Sigma^{RPA}$ is related to the $GW$-approximation as the self-energy in the first self-consistency cycle,[29,37]

$$\Sigma^{RPA}(\tau) = G^0(\tau) W^0(\tau), \quad (11)$$

where $W^0$ in the GW framework is synonymous to $W^{RPA}$. Having the self-energy at hand, one can use Eq. (9) to calculate the RPA density matrix. Note that this RPA density matrix is an approximation to the fully self-consistent GW density matrix[23] in the same way as the linearized GW density matrix.[21,22] The density matrix $n = n^0 + n^{RPA}$ is than diagonalized in the subspace of the virtual states; i.e., the occupied states from the mean field calculation are retained. The occupied orbitals together with the eigenstates obtained from that subspace diagonalization form the RPANOs basis set. Keeping the occupied states has practical reasons. In this way, the RPANOs reproduce the underlying HF or DFT ground state energy exactly, while at the same time the virtual states are rearranged to yield a more compact set for higher hierarchy methods when the orbital set is truncated. RPANOs obtained by this approach therefore provide a versatile basis for different applications.

The RPANOs are sorted by descending occupation number, i.e., eigenvalue with respect to the density matrix, and to obtain a truncated basis with $N$ basis functions, one simply uses the first $N$ RPANOs according to this order. In some cases, it is now necessary to diagonalize the mean-field Hamiltonian in the subspace spanned by the truncated RPANOs basis. For example, to calculate the MP2 energy with 256 RPANOs, the HF Hamiltonian is diagonalized in the subspace spanned by the first 256 RPANOs. The subspace HF orbitals from that diagonalization span the same space as the truncated RPANOs and allow at the same time to use the Brillouin theorem as required by most MP2 implementations.

The computational cost for all the above calculations scales cubically in system size,[40–42] allowing the method to be used for relatively big systems ($\approx$100 atoms) with moderate computing resources ($\approx$100 CPUs).

## B. FCI

The full configuration interaction method (FCI) solves the nonrelativistic Schrödinger equation within a given basis set exactly. Therefore, it provides a useful benchmark for other correlation-consistent methods. Its computational cost scales exponentially with the number of orbitals and the number of electrons, limiting its range of application. In order to obtain the FCI solution, one starts with the HF orbitals and constructs all possible Slater determinants with $N_{occ}$ occupied orbitals. By separating each Slater determinant in its spin up and spin down part, and by using Slater-Condon rules, it is possible to efficiently evaluate contractions of the form $HC$, where $C$ is the vector of Slater determinant weights in the wave function. Following the method of Handy and Knowles[13] and using a Davidson-Liu algorithm to iteratively diagonalize the matrix $H$,[13] the FCI solution can be obtained. Using Eq. (1), the FCI 1-RDM can be easily calculated once the FCI wave function is known. Finally, the diagonalization of this density matrix gives the FCI natural orbitals.

## IV. APPLICATION

All calculations regarding the RPANOs were performed using VASP.[44–46] To obtain the RPANOs, we tried two different approaches, namely, (restricted) HF as well as PBE as underlying single determinant method. However, since the observations for these two approaches were very similar, we will focus on presenting the results that we obtained from using the HF orbitals as the underlying approach for the RPA calculations. For large scale applications on many atom systems, one would most likely though revert to the more efficient DFT approximation. The RPANO calculations are based on the RPA implementation described in previous works.[40–42] For the systems investigated in this work, 8–12 points on the imaginary time/frequency axis were sufficient. For the MP2 calculations in this work, we used the standard routines available in VASP.[47,48] The FCI calculations were performed with a development version, briefly described above. Details can be found in the M.Sc. thesis of Sukurma.[49] Besides using RPANOs, we also performed MP2 and FCI calculations with truncated HF basis sets, for the purpose of comparison. For these truncated HF basis sets, HF orbitals were ordered by ascending HF eigenvalues and only the lowest ones were used. Where not stated otherwise, the shown figures were obtained from calculations with a plane wave cutoff of 750 eV. Note that the orbitals shown below are approximations and do not fulfill the exact cusp condition due to the use of a finite sized plane wave basis set [compare Eq. (10)].

## A. Helium atom

The first benchmark system that we studied is the helium atom. The helium atom is an ideal benchmark system for many body approaches for various reasons. First, unlike other model systems (e.g., jellium), it is a real many body system that can be studied experimentally. Though it is a relatively simple system, its theoretical description faces already prototypical many-body challenges and it is not possible to write down a closed form for its exact solution. Furthermore, a numerically exact solution is available, providing a perfect ground to compare with the results of approximate many body approaches.[50]

The setup was a 10 Å × 10 Å × 10 Å unit cell with periodic boundary conditions. We used a plane-wave basis with different cutoffs up to 750 eV and a $1/r$ potential for the ion.

First, we examined the convergence behavior of the MP2 correlation energy with respect to the RPANOs and compared it with that for canonical HF-orbitals. The results are shown in Fig. 1. While the correlation energy using canonical HF orbitals converges very slowly and would require almost the full basis set (for this setup with a 750 eV plane-wave cutoff, these are $47 \cdot 10^3$ orbitals), with only 128 RPANOs, the correlation energy converged to <1 meV above the exact value. Note that in this case, converged results mean converged for a specific fixed plane wave basis set size. To obtain accurate MP2 correlation energies, one would also need to check convergence with respect to the plane wave cutoff. However, to show the qualitative difference between RPANOs and HF orbitals, it is sufficient to inspect the behavior at a fixed cutoff, where the error in the MP2 energy related to the cutoff is much smaller than the difference between the MP2 energy at a few hundred RPANOs and the same number of canonical HF orbitals.

We repeated the procedure for FCI correlation energies using 10–60 orbitals (see Fig. 2). As was seen for the MP2 correlation energy, canonical HF-orbitals converge the energy very slowly. Again, to reach highly accurate results, one would also need to go to higher plane wave cutoffs, but for a qualitative comparison, the chosen cutoff suffices for the same reasons as at the MP2 level. As a matter of fact, the ground state energy of an FCI calculation with a specific basis is always an upper bound for the real ground state energy. One can conclude from this that the RPANOs are much better in spanning the relevant space of correlated single particle states than the HF-orbitals.

In order to obtain a more accurate approximation to the total energy of the helium atom, $E^{tot} = E^{HF} + E^{cor}$, we first used a plane wave cutoff extrapolation for the FCI correlation and the HF energy.[51,52] The correlation energy was extrapolated from FCI values at plane wave cutoffs of 474 eV, 621 eV, and 750 eV. The HF energy was extrapolated separately from plane wave cutoffs up to 6000 eV in a 16 Å × 16 Å × 16 Å unit cell in order to reach more accurate predictions. This procedure is justified, because the correlation energy is less than 1.5% of the HF energy. For the plane wave extrapolation of the energies, we used a linear regression with respect to the inverse number of plane waves, $N_{PW}^{-1}$, which is itself proportional to $E_{cut}^{-3/2}$,

$$E^{cor/HF}(E_{cut}) \approx E^{cor/HF}(\infty) + p \cdot E_{cut}^{-3/2}. \quad (12)$$

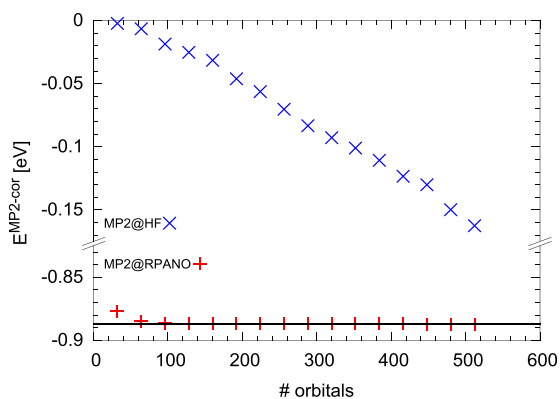

**FIG. 1.** Convergence behavior of the MP2 correlation energy for the He atom. The convergence with respect to the number of RPANOs and HF-orbitals is compared. The black line shows the result for the full basis set. Note that we use two different ranges for the energy.

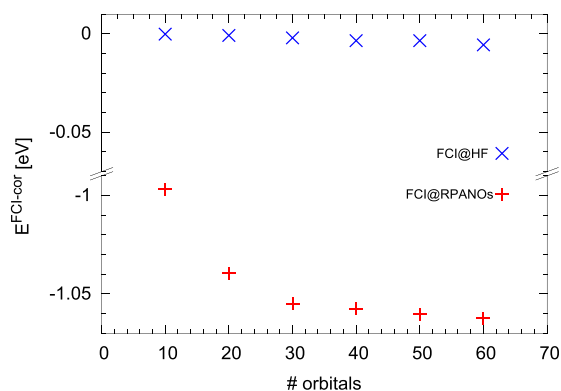

**FIG. 2.** Convergence behavior of the FCI correlation energy for the He atom. The convergence with respect to the number of RPANOs and HF-orbitals is compared. The FCI energy using 10–60 HF virtual orbitals is at least 1 eV above the converged value.

This value is the best approximation we can achieve for a fixed number of truncated RPANOs.

However, the value for the cutoff extrapolated correlation energy still depends on the number of RPANOs used in the FCI calculation. Therefore, to obtain a highly accurate correlation energy, we performed an orbital-set extrapolation on top of the cutoff extrapolation. To achieve this, we calculated cutoff extrapolated FCI correlation energies as described above with different numbers of RPANOs (10, 20, . . . , 60). These cutoff extrapolated values were extrapolated by another linear regression, this time with respect to the inverse number of RPANOs used in the respective FCI calculations,

$$E^{FCI-cor}(N_{RPANOs}) \approx E^{FCI-cor}(\infty) + q \cdot N_{RPANOs}^{-1}. \quad (13)$$

By doing so, we obtained an approximate value for the FCI total energy of −79.019 eV, which differs by less than 1 meV from the literature value from the basis set extrapolation with Gaussian cc-pVxZ basis sets.[50] The extrapolated values for the two methods are therefore well within the error bars of each other. Another "exact" value from Hylleraas-like calculations is −79.014 eV;[53] our estimation is around 5 meV below that. These results are summarized in Fig. 3.

Finally, we compared the shape of the natural orbitals along an axes through the nucleus of the atom. In Figs. 4–6, we show the 1s and 2s orbitals of helium for the different methods we employed. While the occupied 1s orbital is well described in the mean field methods, the 2s orbitals for the different methods vary significantly. This is not very surprising as the HF/DFT states are not computed as eigenstates of the density matrix but stem from a mean field Hamiltonian. In fact, by construction, the virtual HF/DFT states are degenerate eigenstates to the HF/DFT density matrix with eigenvalue 0. This is easily seen by Eqs. (2) and (5), since the mean field single particle states are orthogonal to each other. Furthermore, from Fig. 6, we can deduce that the 2s FCI natural orbital is already converged with as little as 10 RPANOs, but far from converged with 60 HF orbitals. This substantiates the results from Figs. 1 and 2 that canonical HF orbitals calculated from a plane wave basis are impractical for correlated calculations, which is cured by introducing RPANOs.

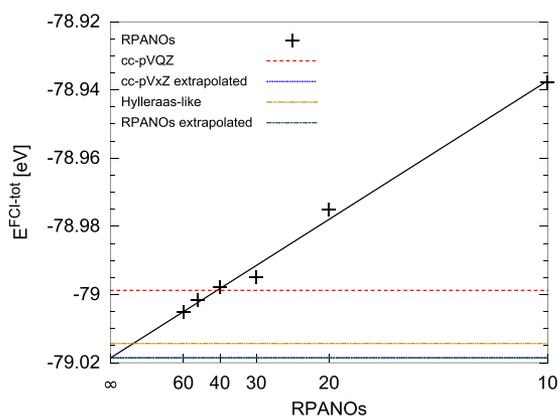

**FIG. 3.** The cutoff extrapolated FCI total energies for the He atom are shown for different numbers of RPANOs. A basis set extrapolation with 1/RPANOs yields a He total energy that is 5 meV below the value from Hylleraas-like calculations. The values for extrapolated RPANOs and extrapolated cc-pVxZ are on top of each other.

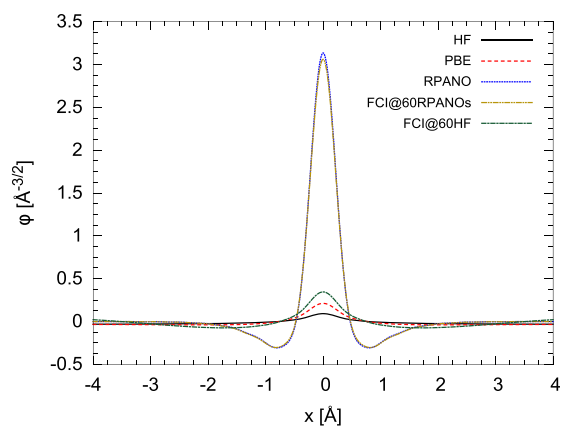

**FIG. 5.** He "2s orbital." The first unoccupied mean field or natural orbital is shown for PBE, HF, RPANO, FCI@60RPANOs, and FCI@60HF.

## B. Hydrogen molecule

The second benchmark system investigated is the hydrogen molecule. Though $H_2$ is the simplest molecule, its exact quantum mechanical description already shows prototypical challenges.

Specifically, the dissociation of the two hydrogen atoms is challenging from a theoretical point of view. In the dissociation limit, an accurate wave function based method should yield a wave function corresponding to two separate hydrogen atoms. The restricted HF method, however, yields an inadequate dissociation limit with an energy substantially above the correct limit due to a flawed mixture of single electron states. Conversely, the spin-unrestricted HF (URHF) method yields a solution with an up electron on one site and a down electron on the other site, which is a spin contaminated singlet. The true singlet would have an equal expectation value for up and down electrons on both sites and is a mixture of (at least) two Slater determinants. The underlying problem is that a single determinant description is insufficient to describe the dissociation process. Similar problems exist in DFT.[54] Therefore, the hydrogen molecule is, despite being seemingly simple, an interesting benchmark system for correlated electronic structure methods.

The computational setup was again a 10 Å × 10 Å × 10 Å unit cell with periodic boundary conditions. We used a plane-wave basis with different cutoffs up to 750 eV. The internuclear distance was varied between the equilibrium distance of 0.74 Å and 5.0 Å.

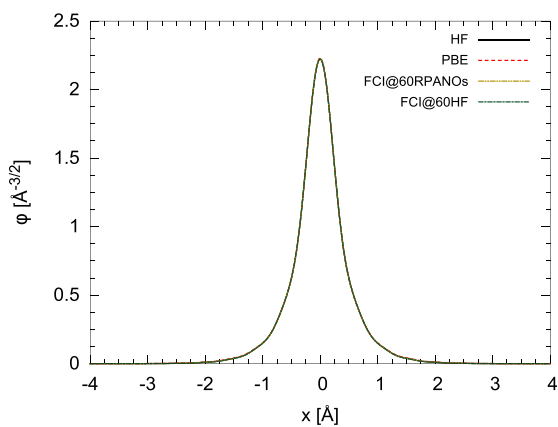

**FIG. 4.** He 1s orbital. The occupied orbital is shown for PBE, HF, FCI@60RPANOs, and FCI@60HF. RPANO and HF are by construction identical.

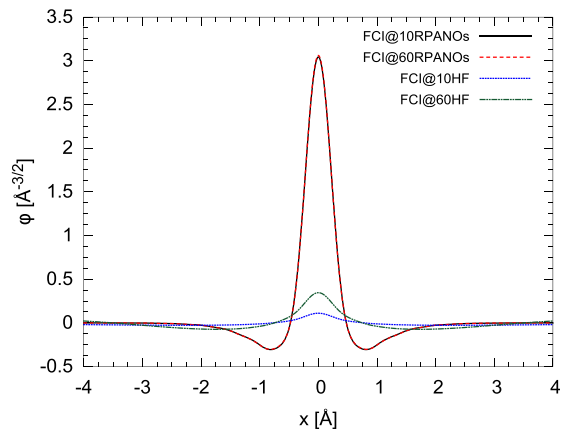

**FIG. 6.** He "2s orbital." The first unoccupied FCI natural orbital is shown for 10 and 60 RPANOs as well as for 10 and 60 HF orbitals.

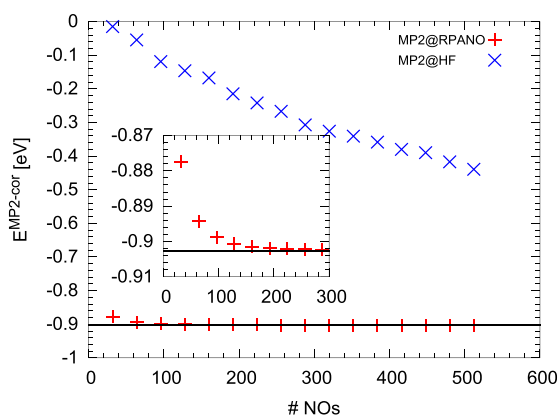

**FIG. 7.** Convergence behavior of the MP2 correlation energy for the $H_2$ molecule at equilibrium distance. The convergence with respect to the number of RPANOs and HF-orbitals is compared. The black line shows the result for the full basis set.

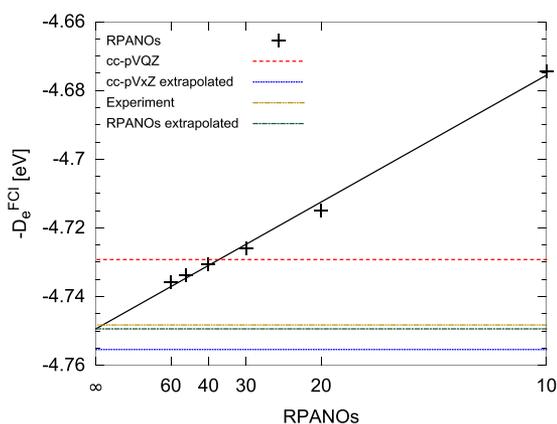

**FIG. 9.** The cutoff extrapolated FCI binding energies for the $H_2$ molecule are shown at different numbers of RPANOs. A basis set extrapolation with 1/RPANOs yields a $H_2$ binding energy that is 1 meV below the corresponding experimental reference value [see Eq. (14)].

As for the helium atom, we examined the convergence behaviors of the MP2 as well as the FCI correlation energy with respect to the RPANOs and compared them with convergence for HF-orbitals. The results of the calculations are shown in Figs. 7 and 8. Similarly to the helium atom, the correlation energies converge very slowly with an increasing number of HF orbitals. Using RPANOs, we only need a few hundred basis functions to obtain converged results for the MP2-energy. The statements on the cutoff convergence in Subsection IV A on the helium atom apply here as well.

For the FCI total energies, a cutoff extrapolation was done in the same fashion as for the helium atom according to Eq. (12) – from cutoffs up to 6000 eV for HF and from cutoffs of 474 eV, 621 eV, and 750 eV for the FCI correlation energy. Additionally, we also performed an orbital set extrapolation with respect to the number of RPANOs [compare Eq. (13)]. In order to compare our results with experimental energies, we calculated the dissociation energy by subtracting twice the URHF energy for the isolated atom obtained with the same procedure (cutoff extrapolation), from the extrapolated FCI total energy at equilibrium distance (0.74 Å). An appropriate experimental value to compare our *ab initio* Born-Oppenheimer approximation result with was obtained by adding the experimental vibrational zero point energy ($E_{ZPE}$)[55] to the experimental dissociation energy $D_0$,[56]

$$D_e = D_0 + E_{ZPE}. \quad (14)$$

Our extrapolated FCI binding energy ($-D_e$) is −4.749 eV, which is 1 meV below the corresponding experimental value of −4.748 eV. Additionally, we performed FCI calculations with the Gaussian pp-VxZ basis sets, with $x = 2, 3, 4$, and extrapolated in order to obtain an estimation for the basis set limit of the $H_2$ binding energy in the Born-Oppenheimer approximation. The extrapolation was done along the lines of Ref. 50; i.e., for the correlation energy, we

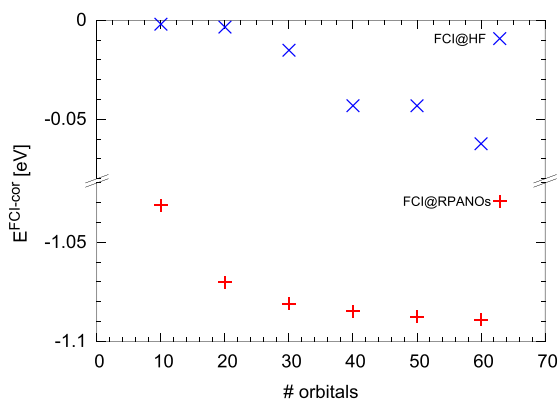

**FIG. 8.** Convergence behavior of the FCI correlation energy for the $H_2$ molecule at equilibrium distance. The convergence with respect to the number of RPANOs and HF-orbitals is compared. The FCI energy using 10–60 HF virtual orbitals is at least 1 eV above the converged value.

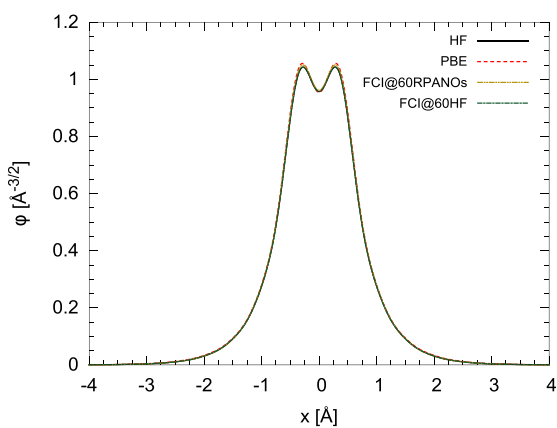

**FIG. 10.** $H_2$ $1\sigma$ orbital at equilibrium distance. The occupied orbital is shown for PBE, HF, FCI@60RPANOs, and FCI@60HF. RPANO and HF are by construction identical.

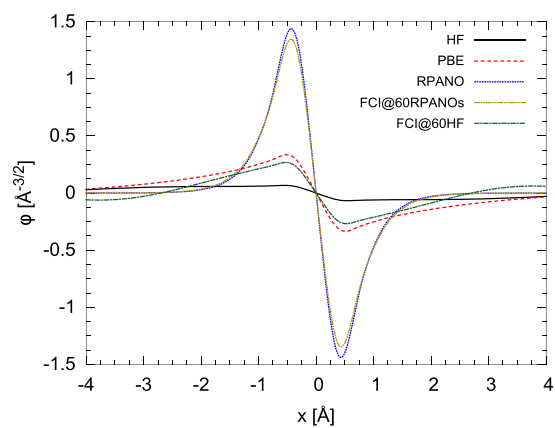

**FIG. 11**. $H_2$ $1\sigma*$ orbital at equilibrium distance. The first unoccupied mean field or natural orbital is shown for PBE, HF, RPANO, FCI@60RPANOs, and FCI@60HF.

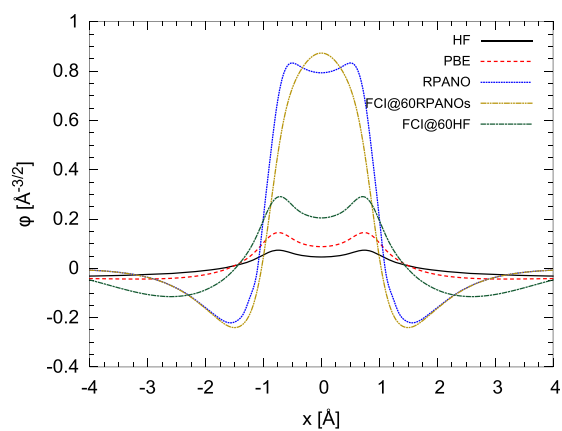

**FIG. 13**. $H_2$ $2\sigma$ orbital at an internuclear distance of 1.588 Å (=3 $a_0$). The second unoccupied mean field or natural orbital ($\hat{=}2\sigma$) is shown for PBE, HF, RPANO, FCI@60RPANOs, and FCI@60HF.

extrapolated according to

$$E^{\text{cor}}(x) = E^{\text{cor}}(\infty) + c \cdot x^{-3}, \tag{15}$$

while for the HF energy, we fitted

$$E^{\text{HF}}(x) = E^{\text{HF}}(\infty) + a \cdot e^{-b \cdot x}. \tag{16}$$

The extrapolated energy from the pp-VxZ basis sets was −4.755 eV, 6 meV below the value for RPANOs and 7 meV below the experimental value. These results are summarized in Fig. 9.

Again, we also inspected the shape of the orbitals, this time along the bond-axis of the hydrogen molecule. In the equilibrium position, the first NO (i.e., the occupied $1\sigma$ state) is well approximated by the mean field methods $1\sigma$ orbital (see Fig. 10). But the second natural orbital is already very different from DFT and even more so from the HF $1\sigma^*$ (see Fig. 11). This has the same reasons as for the He 2s orbital.

In Fig. 11, one can also see that the RPANO does not match the FCINO perfectly. But using just 10 RPANOs for the FCI calculation, the FCI natural orbital is already converged (see Fig. 12) This is a hint that at least in this case, the RPANOs span the relevant space efficiently. In contrast, the HF orbitals are not even close to convergence with 60 orbitals.

Similar observations can be made for other cases, with varying internuclear distances and orbital levels as well as basis set sizes and underlying mean field methods. We have investigated a variety of cases, and we have picked a few illustrative examples in an effort to summarize our findings.

One of these examples is the third orbital ($2\sigma$) at an internuclear distance of 1.588 Å (=3$a_0$). This example is interesting since in this case the RPANO noticeably differs from the FCINO (see Fig. 13). However, inspecting Fig. 14, one can see again that only very few RPANOs are necessary to describe the third FCINO correctly, even though the third RPANO does not match well with the third FCINO. Furthermore, once again, the canonical HF-orbitals poorly describe the FCINO.

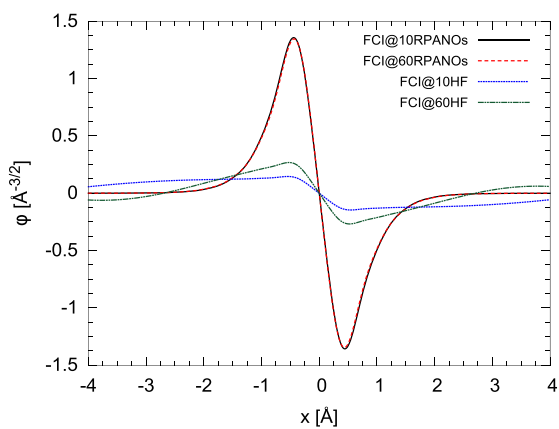

**FIG. 12**. $H_2$ $1\sigma^*$ orbital at equilibrium distance. The first unoccupied FCI natural orbital is shown for 10 and 60 RPANOs as well as for 10 and 60 HF orbitals.

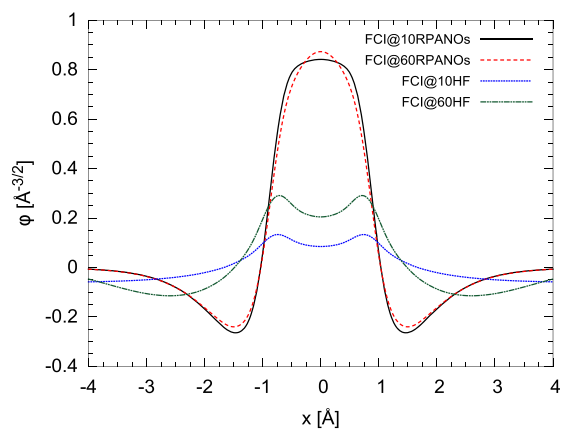

**FIG. 14**. $H_2$ $2\sigma$ orbital at an internuclear distance of 1.588 Å (=3 $a_0$). The second unoccupied FCI natural orbital ($\hat{=}2\sigma$) is shown for 10 and 60 RPANOs as well as for 10 and 60 HF orbitals.

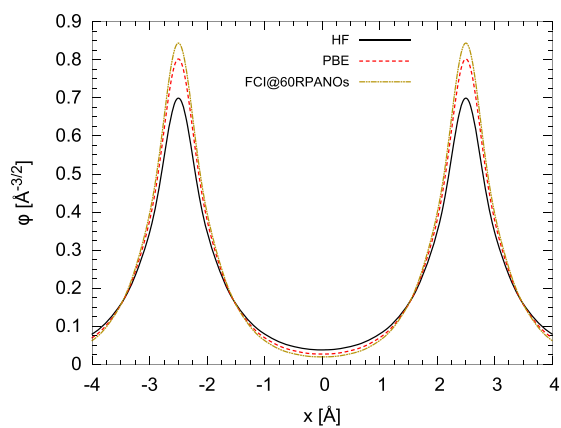

**FIG. 15**. $H_2$ $1\sigma$ orbital at an internuclear distance of 5 Å. The first occupied orbital ($\hat{=}1\sigma$) is shown for PBE, HF, and FCI@60RPANOs.

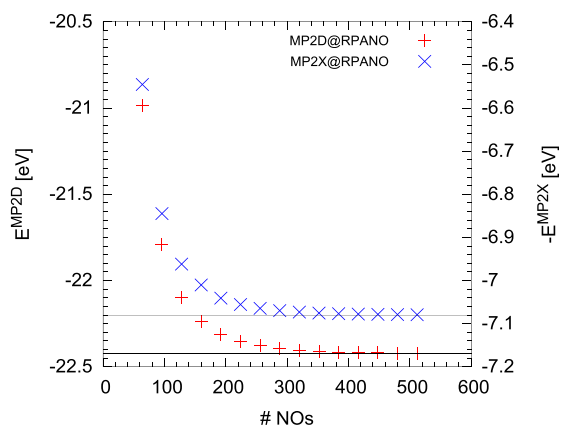

**FIG. 17**. Convergence behavior of the direct and exchange contribution to the MP2 correlation energy for the $F_2$ molecule. The convergence with respect to the number of RPANOs is shown. The black lines indicate the converged result for the full basis set. Note the different scales.

Finally, we show the orbitals at dissociation (5 Å). This time we compare the first two orbitals again. At this distance, the occupied orbitals vary more between the employed methods than they did at equilibrium distance (compare Figs. 10 and 15). By construction, the occupied RPANO is equal to the occupied HF orbitals. It is noteworthy that in this case, the PBE orbitals, and hence the PBE electron density, is much closer to the FCINO than the HF orbital is (Figs. 15 and 16).

### C. Fluorine molecule

In the two electron systems $H_2$ and He, the direct and exchange contribution to the MP2 correlation energy is related by a simple factor of $-2$, i.e., $E_d^{(2)} = -2E_x^{(2)}$. As $\Sigma^{RPA}$ consists of an infinite series of bubble diagrams ($\chi^0$), it intrinsically contains the direct part of the MP2 self-energy $\Sigma_d^{(2)}$ but also accordingly models the exchange part $\Sigma_x^{(2)}$. For these two reasons, we expected the RPANOs to efficiently span the relevant subspace for MP2 calculations in $H_2$ and He. However, for a multielectron system like the fluorine molecule, the simple relation between direct and exchange part of the MP2 energy is no longer valid and it is therefore not clear from the outset that RPANOs will converge the MP2 energy quickly. Thus, we investigated the convergence behavior for the direct (MP2D) and exchange (MP2X) contribution to the MP2 correlation energy with respect to RPANOs separately (Fig. 17). We compare again the performance of RPANOs and HF orbitals for calculating the total MP2 correlation energy (Fig. 18). For these calculations, we employed the PAW method[37,58] and used again a 10 Å × 10 Å × 10 Å unit cell and a plane wave cutoff of 750 eV.

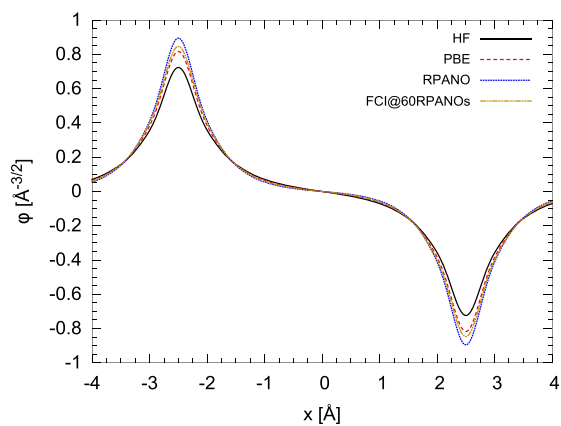

**FIG. 16**. $H_2$ $1\sigma^*$ orbital at an internuclear distance of 5 Å. The first unoccupied mean field or natural orbital ($\hat{=}1\sigma^*$) is shown for PBE, HF, RPANO, and FCI@60RPANOs.

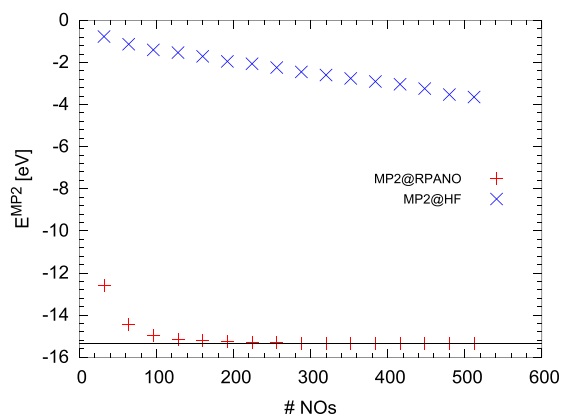

**FIG. 18**. Convergence behavior of the total MP2 correlation energy for the $F_2$ molecule. The convergence with respect to the number of RPANOs and HF orbitals is compared. The black line indicates the converged result for the full basis set.

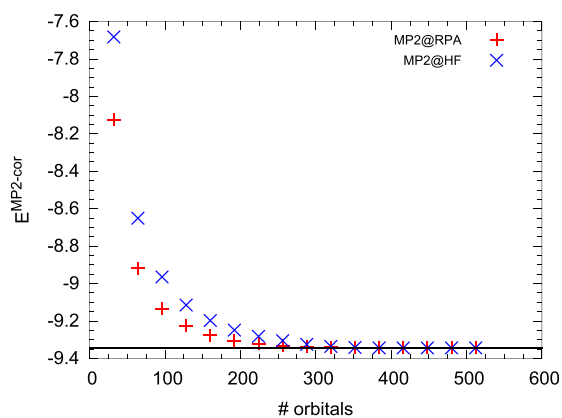

**FIG. 19**. Convergence behavior of the MP2 correlation energy for diamond. The convergence with respect to the number of RPANOs and HF-orbitals is compared. The black line shows the result for the full basis set. The calculation was performed in the primitive cell with experimental lattice parameter $d = 3.567$ Å. We used the PAW method[57,58] with a $4 \times 4 \times 4$ k-mesh and a plane wave cutoff of 600 eV.

We observed that using RPANOs for fluorine, the exchange contribution to the MP2 correlation energy converges rapidly and in the same manner as the direct contribution (see Fig. 17). This is a strong indication that the proposed method, using RPANOs to construct orbitals for more expansive correlated methods, is straightforwardly applicable to complex, multielectron systems. As expected from the previous results, the energy convergence with respect to the number of HF orbitals is again extremely slow compared to the RPANOs (see Fig. 18).

## D. Diamond

The fourth benchmark system, diamond, was chosen to demonstrate the method on a simple periodic multielectron system. Figure 19 shows that the convergence behavior of the MP2 energy is slightly improved by the use of RPANOs, but the effect is much smaller than in the molecular systems. Since we use the primitive cell, the number of plane waves in the full basis set is much smaller than it was for the molecules with large vacuum regions between the periodic images - in this case ≈400 for diamond vs ≈47 000 for $H_2$. In this case, between 200 and 300 RPANOs were sufficient to reach convergence, similar to the numbers for the hydrogen molecule. Thus, the improvement being small is not caused by an underperformance of RPANOs but rather due to a comparatively small number of orbitals in the full basis set.

## V. CONCLUSION

We have tested the efficiency of RPANOs for MP2 and FCI calculations and found that for atoms and molecules they can drastically reduce the number of orbitals that are necessary to reach converged correlation energies in comparison with canonical HF or PBE orbitals. In the present implementation, the calculation of RPANOs scales only cubically with system size. This favorable scaling affords the proposed method an advantage over a similar method using MP2 natural orbitals, which scale with the fifth order of the system size in the canonical implementation (and with fourth order if an approximation is used).[20] In general, the computational cost of correlated methods increases steeply with respect to the number of orbitals, even exponentially for CI. Thus, the computational cost for the preceding compression of the space of orbitals will generally be small compared to the savings gained in the final accurate correlated calculations.

An important finding of the present work is that RPANOs yield a set of orbitals that allows us to converge the correlation energy rapidly, even in cases where the RPA itself is most likely not very accurate. Typical examples for such situations are bond breaking and bond making, in this study exemplified for the case of bond dissociation of $H_2$. Specifically, on dissociation, we found cases where the first few RPANOs deviate significantly from the natural orbitals determined from the FCI correlated density matrix. Regardless of this difficulty, when the correlation energy is expanded in an increasing set of RPANOs, rapid convergence with the number of RPANOs is observed even for these "difficult" cases.

The present method is especially useful for systems composed of large vacuum regions where the basis set size becomes unmanageable when plane waves are used (we are talking of ten thousands of plane wave orbitals). In fact, some of us have already applied a preliminary implementation of the algorithm presented here to calculate RPA natural orbitals and subsequently solve the BSE equations in this smaller basis. In this way, we were able to determine accurate quasiparticle energies in the GWT approximation for molecules.[59] Apart from molecular systems, open structures (zeolites) and surface science studies with large vacuum regions are likely to be an interesting and promising field of application for the RPANO method. Recall that coupled cluster methods scale at least with the fourth order of the number of virtual orbitals. If one is capable to compress the number of virtual orbitals by say a factor 2–3, speed ups of one to two orders of magnitude can be expected. We also note that the RPA is expected to be already fairly accurate for the prediction of adsorption energies;[60,61] however, close to transition states we expect that methods beyond the RPA will be required. Our results for the bond dissociation in $H_2$ give hope that the RPANOs will work also well in such challenging situations. Overall, the present implementation will greatly boost the applicability of plane wave basis sets allowing them to compete with the now omnipresent local Gaussian basis sets used in quantum chemistry.

Last but not least, natural orbitals in periodic systems can be used as a stepping stone for "strongly" correlated calculations, such as dynamical mean field theory (DMFT) or density matrix renormalization group (DMRG) theory. There are already examples in the literature indicating that such an approach might be at least competitive with the usual approach of maximally localized Wannier functions,[62–64] since natural orbitals with an occupancy far from zero or one are likely to contribute most to the correlation energy. Thus, combining perturbative methods for weakly correlated orbitals with an accurate correlation method for the strongly correlated orbitals is an important future development to be pursued.

## ACKNOWLEDGMENTS

Funding by the Austrian Science Fund (FWF): F41 (SFB ViCoM) is gratefully acknowledged.